\documentclass[amssymb,nobibnotes,aps,prd,superscriptaddress,preprintnumbers]{revtex4}
\usepackage{amsmath,amssymb}
\usepackage{color}

\newcommand{\fr}[1]{\frac{1}{#1}}

\newcommand{\ord}[1]{{\mathcal O}(#1)}
\newcommand{\dell}[2]{\frac{\partial #1}{\partial #2}}

\newcommand{\pld}{{p_l}}
\newcommand{\cE}{{\mathcal E}}

\begin{document}

\title{Caged black hole with Maxwell charge}

\author{Ryotaku Suzuki}
\affiliation{Department of Physics, Kyoto University, Kyoto 606-8502, Japan}
\author{Shunichiro Kinoshita}
\affiliation{Osaka City University Advanced Mathematical Institute, Osaka 558-8585, Japan}
\author{Tetsuya Shiromizu}
\affiliation{Department of Physics, Kyoto University, Kyoto 606-8502, Japan}

\date{\today}

\preprint{OCU-PHYS-367, AP-GR-98}

\begin{abstract}
We construct the perturbative solution of the charged black hole
in the Kaluza-Klein spacetime
with the matched asymptotic expansion method.
The corrections to the thermodynamic
 variables are calculated up to the post-Newtonian order.
We confirmed that the method can work very well in the Einstein-Maxwell theory.
\end{abstract}

\maketitle

\section{Introduction}

In the last decade, the higher dimensional black holes have been known 
to have much richer structure than in four
dimensional spacetime~\cite{review}. The discovery of the black ring solution~\cite{ER02,ER03} revealed that
the uniqueness and the horizon topology theorem in four dimension
 are no longer valid in higher dimensions.

The similar breakdown happens in the Kaluza-Klein spacetimes. 
When the mass scale is smaller than the scale of the extra dimension,
there are the black string solutions and the black holes localized in
the Kaluza-Klein spacetime (the caged black hole solutions) 
for the same total mass, which have the different horizon topologies.
Furthermore, there exist the uniform black strings~(UBS) and the non-uniform black strings~(NUBS).
The NUBS phase bifurcates from the UBS phase at the marginally stable mode of the Gregory-Laflamme 
instability~\cite{GL93,GL94}.
This system admits the interesting phenomena such as 
the topology changing phase transition from the NUBS phase to the caged black hole phase,
 and the critical dimension for the stability of the NUBS phase~\cite{BS/BH}.

Since it is difficult to construct the exact solutions of the black objects
with the nontrivial horizon topology,
the numerical or perturbative approach is helpful in the analysis of higher dimensions.
In the Kaluza-Klein spacetime, the small caged black holes have been first studied perturbatively
using a single coordinate patch~\cite{HO02,Har04}.
Gorbonos and Kol used two coordinate patches
 and introduced the systematic procedure to obtain the solutions perturbatively~\cite{GK04}.
This perturbative method is called the matched asymptotic expansion~(MAE).
The MAE can be applied to the spacetime having at least two separate
scales.
Using the MAE, they constructed the small caged black holes up to the post-Newtonian 
order in the asymptotic zone~\cite{GK04,GK05}. 
This method is also useful in the ultra-spinning limit of the black objects~\cite{EHNOR07, EHNO10}.

The effective field theory~(EFT) method~\cite{GR04} is another useful method to systematically obtain the
thermodynamic properties of such spacetime,
which is considered to be equivalent with the MAE method.
In the EFT method, the small scales are integrated out to give the effective Lagrangian
for the coarse-grained black objects.
This is applied to the caged black holes~\cite{CGR06,KS08}.
The caged black holes with charge is also studied by the EFT~\cite{GRS09}.

In this paper, we will generalize the MAE method to the Einstein-Maxwell theory
and construct the perturbative solution of the charged black holes in the Kaluza-Klein spacetime.
In the neutral limit, the solutions here reproduce the neutral results~\cite{GK05}.
We also calculated the correction to the thermodynamic variables
 up to the post-Newtonian order
  which confirms the EFT result~\cite{GRS09}.

The organization of this paper is as follows.
In Sec.~II, we explain the setup of the caged black hole and the
matching method. 
In Sec.~III, the Newtonian order is solved at
the asymptotic zone. 
In Sec.~IV, we solve the near zone perturbation.
In Sec.~V, we match the near zone results with the results in Sec.~III
and evaluate some of the thermodynamic variables.
In Sec.~VI, we solve the
post-Newtonian equation at the asymptotic zone.
In Sec.~VII, we compute the PN correction to the global charges,
and analyze the thermodynamic properties of solutions.
We summarize our work in Sec.~VIII. In Appendices, we will present some useful 
formulae and the details of the computations.

\section{Matched Asymptotic Expansion}\label{MAE-explain}

In this section, we introduce the method of the matched asymptotic expansion
in the $(n+3)$-dimensional Kaluza-Klein spacetime
with the Einstein-Maxwell theory.
The action is given by 
\begin{eqnarray}
S =\int \left(\fr{16\pi G}R - \fr{4}F_{\mu\nu}F^{\mu\nu}\right)\sqrt{-g}d^{n+3}x,
\end{eqnarray}
where $F_{\mu\nu}$ is the field strength of the Maxwell field $A_\mu$
defined by $F_{\mu\nu} = \partial_\mu A_\nu - \partial_\nu A_\mu$.
In this paper we choose the unit of  $G =1$, hereafter. 
The Einstein equation becomes
\begin{eqnarray}
 R_{\mu\nu}-\fr{2}Rg_{\mu\nu}= 8\pi T_{\mu\nu},
\end{eqnarray}
where the energy-momentum tensor is given by
\begin{eqnarray}
 T_{\mu\nu}=F_\mu{}^{\lambda}F_{\nu\lambda}-\fr{4}F_{\alpha\beta}F^{\alpha\beta}g_{\mu\nu},
\end{eqnarray}
and the field equation for the Maxwell field is 
\begin{eqnarray}
 F^{\mu\nu}{}_{;\mu} = 0.
\end{eqnarray}
We consider a static localized black hole solution 
in the Kaluza-Klein spacetime which is called ``caged black hole''.
Then we take the small black hole limit
in which the scale of the black hole $\rho_0$
is much smaller than the compactification scale $L$.
Since the Maxwell charge is bounded above by
the extreme limit, it can be assumed to be the same order as the mass
$\sim \rho_0^{n}$.
Therefore, we can perform the similar treatment with the neutral case~\cite{GK04,GK05}.

In this small black hole limit, we have two asymptotic zones.
The spacetime asymptotes to the flat Kaluza-Klein spacetime
in the asymptotic zone~($\rho \gg \rho_0$),
and the charged spherical black hole
in the near zone~($\rho \ll L$).
Let us introduce two coordinate systems.
In the asymptotic zone, it is better to use the cylindrical coordinates
$(r,z)$ as 
\begin{eqnarray}
 ds^2 = -dt^2 + dz^2 + dr^2 + r^2 d\Omega_{n}^2 \label{KK-metric}
\end{eqnarray}
in which the $z$-direction is periodic as $z \sim z+L$.
The periodicity requires that we must consider the gravity from the mirror
images of the black hole. This is equivalent to the infinite array of
black holes in the higher dimensional Minkowski spacetime.
In the near zone, on the other hand, 
it is better to use the spherical coordinates $(\rho,\chi)$ in order to see the
black hole perturbation as 
\begin{eqnarray}
 ds^2 = -dt^2 + d\rho^2 + \rho^2 (d\chi^2 + \sin^2\chi d\Omega_{n}^2).
\end{eqnarray}
In the asymptotic zone, the relation between two coordinate systems is
\begin{eqnarray}
 r=\rho \sin\chi, \;
 z=\rho \cos\chi.
\end{eqnarray}

We expand the metric and the Maxwell field in both zones.
In the near zone, we consider the black hole perturbation with the expansion parameter $1/L $ as
\begin{eqnarray}
 g_{\mu\nu}^{({\rm near})}
= g^{({\rm BH})}_{\mu\nu} + \sum_{k=1}^\infty h_{\mu\nu}^{({\rm near},k)},
\end{eqnarray}
where $h^{({\rm near},k)} \sim \ord{1/L^k}$.
$g^{({\rm BH})}_{\mu\nu}$ is the metric of
the $(n+3)$-dimensional background charged black hole spacetime
\begin{eqnarray}
 g^{({\rm BH})}_{\mu\nu}dx^\mu dx^\nu &=& - f(\rho)dt^2 + \fr{f(\rho)}d\rho^2 +
  \rho^2 d\Omega_{n+1}^2, \label{bg-blackhole}
\end{eqnarray}
where
\begin{eqnarray}
 f(\rho) &=& 1 - \frac{\rho_0^{n}}{\rho^{n}} +
  \frac{\sigma^2}{4}\frac{\rho_0^{2n}}{\rho^{2n}}, \label{bg-blackhole-f}
\end{eqnarray}
and $\sigma$ is a dimensionless charge parameter which becomes $|\sigma|=1$
in the extreme limit.
The total mass of this black hole is
\begin{eqnarray}
 M_0  = \frac{(n+1)\omega_{n+1}\rho_0^n}{16\pi},
\end{eqnarray}
where $\omega_{k} = 2\pi^\frac{k+1}{2}/\Gamma(\frac{k+1}{2})$ is the
area of the $k$-dimensional unit sphere, $S^k$.
The background Maxwell field is 
\begin{equation}
 A_\mu dx^\mu = \frac{Q_0}{n\omega_{n+1}\rho^n}dt,
\end{equation}
where the total charge of the background black hole is given by 
\begin{equation}
 Q_0 = \int * F = \sqrt{\frac{n(n+1)}{32\pi}}\omega_{n+1}\rho_0^n \sigma 
\end{equation}
and non-vanishing component of the field strength is the electric field
\begin{equation}
 F_{t\rho} = \sqrt{\frac{n(n+1)}{32\pi}}\sigma \frac{\rho_0^n}{\rho^{n+1}}
  \equiv \cE . \label{gauge-field-bg}
\end{equation}

In the asymptotic zone, we consider the post-Newtonian expansion
with the expansion parameter $\rho_0$ as
\begin{eqnarray}
 g_{\mu\nu}^{({\rm asym})}
 = \eta_{\mu\nu} + \sum_{m=1}^\infty h_{\mu\nu}^{({\rm asym},m)},
\end{eqnarray}
where $h^{({\rm asym},m)} \sim \ord{\rho_0^m}$ and $\eta_{\mu\nu}$ is
the metric of
the $(n+3)$-dimensional Kaluza-Klein spacetime (\ref{KK-metric}).
The Maxwell field is also expanded in the same way.
It is worth noting that the Newtonian approximation is the order of
$\rho^n_0$ because the mass $M$ is written as $\rho^n_0 \sim G M$.

Now, it is ready to consider the matching procedure between the two zones.
In the limit $\rho_0 \ll L$, there exists the overlap region~($\rho_0 \ll \rho \ll L$ ),
in which we will match the two expansions order by order.
The ``matching ladder'' can be understood by the dimensional counting.
If we take the harmonic gauge, the linear part of
the post-Newtonian equation in the order of $\rho_0^m$ becomes 
\begin{eqnarray}
 \Box h_{\mu\nu}^{({\rm asym},m)} = 0, \label{laplace-eq}
\end{eqnarray}
where $\Box = \eta^{\mu\nu}\partial_\mu \partial_\nu$.
Since we consider the static spacetime, it reduces to the Laplace equation.
The homogeneous solutions with a multipole $l$ behave as $\rho^l$ and $\rho^{-l-n}$.
For the asymptotic zone, the near zone black hole seems to have multipole moments and behave as
 $\rho^{-l-n}$.
From the dimensional counting, 
the terms with $\rho^{-l-n}$ 
in
asymptotic solution of $\ord{\rho_0^m}$ appears in the dimensionless form
\begin{eqnarray}
 h_{\mu\nu}^{({\rm asym},m)} \ni \frac{\rho_0^m }{L^{m-l-n}\rho^{l+n}} .
\end{eqnarray}
Therefore, the near solution of $\ord{1/L^k}$ contributes
to the asymptotic solution with the multipole $l$ of $\ord{\rho_0^{k+n+l}}$.

Similarly, 
the gravity from the mirror images of the black hole affects the near zone geometry
in the form of multipole moments at  the infinity, which are proportional to $\rho^l$.
The dimensional counting shows
\begin{eqnarray}
 h_{\mu\nu}^{({\rm near},k)} \ni \frac{\rho_0^{k-l} \rho^l}{L^k}.
\end{eqnarray}
Then, the asymptotic solution of $\ord{\rho_0^m}$ determines to
the near solution with the multipole $l$ of $\ord{1/L^{m+l}}$.

\section{Newtonian potential in asymptotic zone}

We start the matching of the near zone solutions with the Newtonian order solution in the asymptotic zone.
We will omit the script ``near" or ``asym" for the brevity.
As explained in the previous section, the near zone background metric of $\ord{1/L^0}$ gives
the correction to the asymptotic metric in $\ord{\rho_0^n}$.
At this order, the solution is just the Newtonian potential from the
array of the point source as in the neutral case \cite{GK04}.
In the harmonic gauge 
$\partial_\mu \bar{h}^{\mu\nu}=0$
where $\bar{h}_{\mu\nu}=h_{\mu\nu}-\fr{2}h  \eta_{\mu\nu}$, the linearized
Einstein equation is
\begin{eqnarray}
 -\fr{2}\Box \bar{h}_{\mu\nu}^{(n)} = 8\pi T_{\mu\nu}^{(n)}. \label{newton-asym-einstein}
\end{eqnarray}

Since $T_{\mu\nu}$ is written in terms of the square of $F_{\mu\nu}$ and 
the leading order of $F_{\mu\nu}$ is the Newtonian order,
the Einstein equation is the same as the vacuum case at this order 
(namely, $T_{\mu\nu}^{(n)}=0$).
The boundary condition in the overlap region is determined by the
behavior of the near background metric.
In the overlap region, the near zone background metric in the harmonic coordinates
is given as $\ord{1/L^0}$ terms in Eq.~(\ref{gauge-transformed}), which becomes
\begin{subequations}\label{Newtonian-bc}
\begin{eqnarray}
 g^{({\rm BH})}_{tt} &\simeq& -1 + \frac{\rho_0^n}{\rho^n}, \label{Newtonian-bc1}\\
 g^{({\rm BH})}_{ij} &\simeq& \left(1 + \frac{\rho_0^n}{n\rho^n}\right)\delta_{ij}. \label{Newtonian-bc2}
\end{eqnarray}
\end{subequations}
Requiring the periodicity in $z$-direction and the harmonic condition, the homogeneous solution
with a pole at the origin is given by
\begin{subequations}\label{Newtonian-grav}
\begin{eqnarray}
 h_{tt}^{(n)}&=& \Phi \equiv \rho_0^n \sum_{k=-\infty}^\infty \fr{(r^2+(z-kL)^2)^{\frac{n}{2}}},\label{Newtonian-grav1}\\
 h_{ij}^{(n)}&=& \fr{n}\Phi \delta_{ij}. \label{Newtonian-grav2}
\end{eqnarray}
\end{subequations}
Since $\Phi \simeq \rho_0^n/\rho^n$ as $\rho \ll L$, the coefficients are determined by the match with Eq.~(\ref{Newtonian-bc}).
This is equivalent to the Newtonian potential from the infinite array of point
masses,
\begin{eqnarray}
 T_{tt}^{(n)} &=& \sum_{k=-\infty}^\infty M_0 \delta^{n+1}({\bf x})\delta(z-kL).
\end{eqnarray}
Imposing the Lorenz gauge $\partial^\mu A_\mu^{(n)} =0$, the Maxwell equation is
\begin{eqnarray}
 \Box A_{\mu}^{(n)} = 0.
\end{eqnarray}
In the same way, the solution is given by
\begin{eqnarray}
 A^{(n)}_t = \phi &\equiv& \frac{Q_0}{n\omega_{n+1}} \sum_{k=-\infty}^\infty \fr{(r^2+(z-kL)^2)^{\frac{n}{2}}}\label{Newtonian-maxwell}. 
\end{eqnarray}
This is also equivalent to the potential obtained from the infinite array of
point charges,
\begin{eqnarray}
 J_t^{(n)} = \sum^\infty_{k=-\infty} Q_0\delta^{n+1}({\bf x})\delta(z-kL).
\end{eqnarray}
Now we can see that the gauge potential is proportional to the Newtonian potential,
\begin{eqnarray}
 \phi = \frac{Q_0}{n\omega_{n+1}\rho_0^n}\Phi = \sqrt{\frac{n+1}{32\pi n}}\sigma\Phi. \label{gauge-to-newton}
\end{eqnarray}

$\Phi$ is expanded in the overlap region for $r,z \ll L$,
\begin{eqnarray}
 \Phi = \frac{\rho_0^n}{\rho^n} + \sum_{k=0}^\infty 2\zeta(n+2k)\frac{\rho_0^n\rho^{2k}}{L^{n+2k}}C_{2k}^\frac{n}{2}(\cos\chi) ,\label{newton-expand}
\end{eqnarray}
where $C_{l}^\nu(x)$ is  the Gegenbauer polynomial (see Appendix~\ref{app:Gegenbauer} for the definition).

\section{Near zone perturbation}

We now proceed to the leading order in the near zone.
The leading order correction in the near zone comes from
the asymptotic solution of $\ord{\rho_0^n}$ in the previous section.
In this paper, we only consider the monopole $(l=0)$ correction which is $\ord{1/L^n}$.

Here, we calculate the linear perturbation around the black hole metric of Eq. 
(\ref{bg-blackhole}) as $g_{\mu\nu} = g^{({\rm BH})}_{\mu\nu} + h_{\mu\nu}$.
Since we consider the static perturbation, we can set the following ansatz for the metric perturbation
\begin{eqnarray}
 ds^2 = - f(1+A_0)dt^2 &+& f^{-1}(1 + B_0)d\rho^2+ \rho^2 (1 + E_0)d\Omega_{n+1}^2, \label{monopole-ansatz-metric}
\end{eqnarray}
and for the Maxwell field perturbation,
\begin{eqnarray}
 A_\mu dx^\mu = \left(\frac{Q_0}{n\omega_{n+1}\rho^n} + a_0 \right)dt.\label{monopole-ansatz-field}
\end{eqnarray}
We write the linearized Einstein equation as
\begin{eqnarray}
 \delta R_{\mu\nu} = 8\pi\left[\delta T_{\mu\nu}
 - \fr{n+1}(g^{(\mathrm{BH})\alpha\beta}\delta T_{\alpha\beta} -  T^{(0)}{}^{\alpha\beta}h_{\alpha\beta})g^{(\mathrm{BH})}_{\mu\nu}
-\fr{n+1}T^{(0)} h_{\mu\nu}\right] \equiv 8\pi \delta S_{\mu\nu},\label{einstein1}
\end{eqnarray}
where $T_{\mu\nu}^{(0)}$ is the energy-momentum tensor of the background. 
The components of $T^{(0)}_{\mu\nu}$  are given by
\begin{eqnarray}
 -T^{(0)}{}^t{}_t = -T^{(0)}{}^r{}_r =  T^{(0)}{}^\chi{}_\chi =  T^{(0)}{}^{\theta_i}{}_{\theta_i} =  
  \fr{2}\cE^2 =  \frac{n(n+1)\sigma^2}{64\pi}\frac{\rho_0^{2n}}{\rho^{2n+2}},
\end{eqnarray}
where $\{\theta_i\}_{i=1..n}$ are coordinates of $S^n$. The explicit forms for each components of 
$\delta R_{\mu\nu}$ and $\delta S_{\mu\nu}$ are given in Eqs.~(\ref{multi-ricci}) and (\ref{s-tensor}).
The Maxwell equation becomes
\begin{eqnarray}
 \partial_\rho [\rho^{n+1} a_0' - \frac{Q_0}{2\omega_{n+1}}((n+1)E_0 - A_0 - B_0)]=0.
\end{eqnarray}
This can be easily integrated as
\begin{eqnarray}
 a_0' = \frac{\cE}{2}((n+1)E_0 - A_0 - B_0 + 2\alpha_0), \label{monopole-apsol}
\end{eqnarray}
where $\alpha_0$ is an integral constant.
Using the above and Eq.~(\ref{s-tensor}), $\delta S_{\mu\nu}$ becomes
\begin{subequations}\label{emtensor1b}
\begin{eqnarray}
  \delta S{}^t{}_{t} &=& \frac{n^2\sigma^2}{4}
  \frac{\rho_0^{2n}}{\rho^{2n+2}}[(n+1)E_0 - A_0 + 2\alpha_0], 
  \label{emtensor1b-tt}\\
  \delta S{}^\rho{}_{\rho} &=& \frac{n\sigma^2}{4}
  \frac{\rho_0^{2n}}{\rho^{2n+2}}[(n+1)E_0 - B_0 + 2\alpha_0] 
  \label{emtensor1b-rhorho}
  \end{eqnarray}
  and
  \begin{eqnarray}
  \delta S {}^\chi{}_\chi &=& - \frac{n\sigma^2}{4}
  \frac{\rho_0^{2n}}{\rho^{2n+2}}[E_0 -A_0-B_0+2\alpha_0],\label{emtensor1b-chichi}
\end{eqnarray}\end{subequations}
where $\delta S^\mu{}_\nu = g^{(\mathrm{BH})\mu\alpha}\delta S_{\alpha\nu}$.
Subtracting the $\rho\rho$-component from the $tt$-component
in Eq.~(\ref{einstein1}),
we obtain the following equation 
\begin{eqnarray}
 \rho E_0'' + 2E'_0 - A_0' -B_0' = 0 ,\label{monopole-eq1}
\end{eqnarray}
which is the same with the neutral cases.
Another independent equation comes from the $\chi\chi$-component in Eq.~(\ref{einstein1}),
\begin{eqnarray}
 &&\fr{2}\rho^2(fE_0'' + f'E_0')+ (n+1)\rho fE_0' + \fr{2}\rho f (A_0' -  B_0') \nonumber\\ 
  && + n(E_0-B_0)= \frac{n\sigma^2}{4}\frac{\rho_0^{2n}}{\rho^{2n}}[(n+1)E_0 - B_0 +
  2\alpha_0].
\label{monopole-eq-chichi}
\end{eqnarray}
Integrating Eq.~(\ref{monopole-eq1}), we have 
\begin{eqnarray}
 \rho E_0' + E_0 - A_0 - B_0 = -2C_1 ,\label{monopole-eq1-int}
\end{eqnarray}
where $C_1$ is an integral constant. Imposing 
\begin{eqnarray}
 A_0+B_0+(n-1)E_0 = 0 \label{monopole-res-gauge}
\end{eqnarray}
as the residual gauge condition, Eq.~(\ref{monopole-eq1-int}) can be solved as 
\begin{eqnarray}
 E_0 =  \frac{2C_1}{n} + \frac{C_2\rho_0^n}{\rho^n},\label{monopole-sol-1}
\end{eqnarray}
where $C_2$ is an integral constant.

To solve Eq.~(\ref{monopole-eq-chichi}) in the gauge condition
(\ref{monopole-res-gauge}), we introduce a new variable
defined by
\begin{eqnarray}
 \Psi \equiv -\fr{n-1}(A_0 + nB_0) =E_0 - B_0 ,
\end{eqnarray}
where $A_0$ and $B_0$ are written by $\Psi$ as
\begin{eqnarray}
 A_0 = \Psi - nE_0,\;
 B_0 = E_0  - \Psi .\label{monopole-ABsol}
\end{eqnarray}
Using the above and Eq.~(\ref{monopole-sol-1}), Eq.~(\ref{monopole-eq-chichi}) becomes
\begin{eqnarray}
 \rho (f\Psi)' + nf\Psi =
  \frac{n\rho_0^{2n}}{2\rho^{2n}}
     [\sigma^2(\alpha_0+C_1) + nC_2]
\end{eqnarray}
and then the solution is given by 
\begin{eqnarray}
 f\Psi = \frac{\tilde{C_3}\rho_0^n}{\rho^n} - \frac{\rho_0^{2n}}{2\rho^{2n}}
[\sigma^2(\alpha_0 + C_1) + n C_2] \label{monopole-sol-2},
\end{eqnarray}
where $\tilde{C_3}$ is an integral constant. 
Even after imposing the gauge condition~(\ref{monopole-res-gauge}), there is still remaining gauge degree of freedom $\rho \rightarrow \rho + \delta C\rho_0^n/2\rho^{n-1}$ which changes Eqs.~(\ref{monopole-sol-1}) and (\ref{monopole-sol-2}) as
\begin{subequations}\label{res-gauge-trans}
\begin{eqnarray}
  E_0 &\rightarrow& \frac{2C_1}{n} + \frac{C_2'\rho_0^n}{\rho^n}, \label{res-gauge-trans-e}\\
   f\Psi &\rightarrow& \frac{\tilde{C_3}' \rho_0^n}{\rho^n}  - \frac{\rho_0^{2n}}{2\rho^{2n}}
[\sigma^2(\alpha_0 + C_1) + n C_2'].\label{res-gauge-trans-psi}
\end{eqnarray}\end{subequations}
where $C_2' = C_2 + \delta C$ and $\tilde{C_3}' = \tilde{C_3} + n\delta C$.
Therefore $C_2$ can be the pure gauge and we define
the gauge invariant combination $C_3 \equiv \tilde{C_3} - nC_2$.
Here, $C_3$ is the ambiguity of the bare mass and $\alpha_0$ is the ambiguity of the bare charge
which we can freely choose.

As the boundary condition at the horizon, we impose the regularity 
conditions for $A_0$, $B_0$ and $E_0$
which fix the horizon position.
We write the position of the outer event horizon as $\rho_+$, which is
the larger root of $f(\rho)=0$, 
\begin{eqnarray}
 \rho_+^n =  \fr{2}\rho_0^n(1 + \sqrt{1-\sigma^2}).
\end{eqnarray}
The regularity of $\Psi$ at $\rho=\rho_+$ is guranteed, this condition is satisfied if 
the right-hand side of Eq.~(\ref{monopole-sol-2}) vanishes as $\rho \rightarrow \rho_+$, which gives
\begin{eqnarray}
 \fr{2}\sigma^2(\alpha_0+C_1) + \fr{2}nC_2 = (C_3 + nC_2)\frac{\rho_+^n}{\rho_0^n} \label{monopole-regularity-raw}.
\end{eqnarray}
In general, the perturbed spacetime may not become extremal or neutral when the background spacetime is extremal
($\sigma = 1$) or neutral ($\sigma = 0$).
Then, we will further fix the parameter region of $\sigma$ so that the solution becomes extremal
when $\sigma = 1$ and neutral when $\sigma = 0$.
The extremal condition is attained by imposing that the right-hand side of Eq.~(\ref{monopole-sol-2}) 
has the double roots, $(f\Psi)'|_{\rho=\rho_+, \sigma=1} = 0$, which becomes 
\begin{eqnarray}
 (\alpha_0+C_1 +nC_2)|_{\sigma = 1} = 0 = (C_3 + nC_2)|_{\sigma =1}  \label{monopole-extreme},
\end{eqnarray}
where we used Eq.~(\ref{monopole-regularity-raw}).
The neutral condition is
\begin{eqnarray}
  (2C_3 + nC_2)|_{\sigma=0} = 0.
\end{eqnarray}
Although $C_i$ and $\alpha_0$ may depend on $\sigma$ in general, 
we can set these to satisfy the above conditions for not only 
$\sigma=0, 1$ but also arbitrary $\sigma$ as 
\begin{eqnarray}
 C_2 = C_3 = 0, \; \alpha_0 = - C_1.\label{fix-gauge-ambiguity}
\end{eqnarray}
Therefore, $\Psi$ becomes to be zero, that is, 
\begin{eqnarray}
\Psi = 0.
\end{eqnarray}
Then, from Eqs.~(\ref{monopole-apsol}), (\ref{monopole-sol-1}) and (\ref{monopole-ABsol}),
the perturbation becomes 
\begin{subequations}\label{1stsols}
\begin{eqnarray}
 A_0 &=& -2C_1, \label{1stsols-A}\\
 B_0 &=& \frac{2C_1}{n},  \label{1stsols-B}\\
 E_0 &=& \frac{2C_1}{n} ,\label{1stsols-E}\\
 a_0 &=& - \frac{C_1Q_0}{n\omega_{n+1}\rho^{n}} + \alpha_1 ,
\label{1stsols-a}
\end{eqnarray}\end{subequations}
where $\alpha_1$ is an integral constant.
As a result, the near solutions depend on two parameters $C_1$ and $\alpha_1$,
which are determined by matching with the asymptotic solutions in the
overlap region.

\section{Matching from the Newtonian potential}

As mentioned in Sec.~\ref{MAE-explain}, the near zone monopole correction of $\ord{1/L^n}$
comes from the monopole parts of the Newtonian order asymptotic solution. 
The monopole from the asymptotic zone behaves as a constant which has the dimensionless form, $\rho_0^n/L^n$ at the Newtonian order.

Since the leading order terms in the near zone perturbation
are not affected by the gauge transformation into the harmonic coordinates (\ref{gauge-trans}), the
near solution in the overlap region $(\rho_0 \ll \rho)$ becomes
\begin{subequations}\label{1stmatch}
\begin{eqnarray}
 h_{tt}^{(\mathrm{near},n)}&=&-fA_0   \simeq 2C_1,\label{1stmactch-t}\\
 h_{\rho\rho}^{(\mathrm{near},n)}&=& \frac{B_0}{f} \simeq \frac{2C_1}{n},\label{1stmatch-r}\\
 h_{\chi\chi}^{(\mathrm{near},n)}&=& \rho^2E_0 = \frac{2C_1}{n}\rho^2.\label{1stmatch-c}
\end{eqnarray}\end{subequations}
On the other hand, the $\rho_0^n/L^n$ term in the expansion of the 
Newtonian potential (\ref{newton-expand}) is 
\begin{eqnarray}
 h^{({\rm asym}, n)}_{tt}= \Phi = \frac{\rho_0^n}{\rho^n} + 2\zeta(n) \frac{\rho_0^n}{L^n} + \cdots
\end{eqnarray}
and then the matching provides
\begin{eqnarray}
 C_1 = \zeta(n)\frac{\rho_0^n}{L^n}.
\end{eqnarray}

We now define the expansion parameter $\lambda = \zeta(n)\rho_0^n/L^n$. 
Also, for the Maxwell field the leading order term of the near solution
matches the gauge potential $\phi$ of the asymptotic solution in the overlap region as 
\begin{equation}
 A^{(\mathrm{near},n)}_t \simeq \alpha_1
 = \frac{2Q_0}{n\omega_{n+1}\rho_0^n}\lambda.
\end{equation}

Then, combining Eq.~(\ref{1stsols})
with the ansatz~(\ref{monopole-ansatz-metric}) and (\ref{monopole-ansatz-field})
, the near zone metric up to $\ord{1/L^n}$ becomes
\begin{subequations}\label{1stsol-matched}
\begin{eqnarray}
 g_{tt}^{(\mathrm{near})} &=& -f( 1 - 2\lambda) + \ord{1/L^{2n}} ,
 \label{1stsol-matched-t}\\
 g_{\rho\rho}^{(\mathrm{near})} &=& \fr{f}\left(1 + \frac{2\lambda}{n}\right) + \ord{1/L^{2n}},
 \label{1stsol-matched-r}\\
 g_{\chi\chi}^{(\mathrm{near})} &=& \rho^2 \left( 1 + \frac{2\lambda}{n}\right)+ \ord{1/L^{2n}},
\label{1stsol-matched-c}
\end{eqnarray}\end{subequations}
and the gauge potential in the near zone is 
\begin{eqnarray}
 A_t^{(\mathrm{near})} = \frac{Q_0}{n\omega_{n+1}\rho^n} + \frac{2\lambda Q_0}{n\omega_{n+1}\rho_0^n}\left(1 - \frac{\rho_0^n}{2\rho^n}\right)+ \ord{1/L^{2n}}.\label{1stsol-matched-A}
\end{eqnarray}
From the leading order near solution (\ref{1stsol-matched}), the
correction to the local constants can be computed.
The surface gravity $\kappa$ becomes
\begin{eqnarray}
 \kappa &=& \left. \fr{2}\frac{|\partial_\rho
	     g_{tt}|}{\sqrt{-g_{tt}g_{\rho\rho}}}\right|_{\rho=\rho_+}\nonumber\\
&=&\left.\fr{2}|f'|\left( 1 + \frac{f}{f'}A_0' + \fr{2}(A_0 -
		B_0)\right)\right|_{\rho = \rho_+}\nonumber\\
&=& \frac{n\rho_0^n}{2\rho_+^{n+1}}\sqrt{1-\sigma^2}\left(1- \frac{n+1}{n}\lambda\right) \nonumber\\
&=& \kappa_0\left(1-\frac{n+1}{n}\lambda\right),\label{surface-gravity}
\end{eqnarray}
where $\kappa_0 = \sqrt{1-\sigma^2}n\rho_0^n/2\rho_+^{n+1}$ is the surface gravity for $L=\infty$.
The horizon area $\mathcal A$ becomes
\begin{eqnarray}
 \mathcal{A} &=& \omega_{n+1}(g_{\chi\chi})^\frac{n+1}{2}|_{\rho=\rho_+}\nonumber\\
 &=& \omega_{n+1}\rho_+^{n+1}\left( 1 + \frac{n+1}{n} \lambda \right)\nonumber\\
  &=& \mathcal{A}_0 \left( 1 + \frac{n+1}{n} \lambda \right),\label{area}
\end{eqnarray}
where $\mathcal{A}_0$ is the horizon area for $L=\infty$.

We can also calculate the electrostatic potential $U$,
\begin{eqnarray}
 U \equiv A_t|_{\rho=\rho_+} - A_t|_{\rho=\infty} &=&
 \frac{Q_0}{n\omega_{n+1}\rho_+^n} (1 +\sqrt{1-\sigma^2}\lambda )\nonumber\\
   &=& U_0( 1 + \sqrt{1-\sigma^2}\lambda) ,\label{electrostatic-potential}
\end{eqnarray}
where $U_0$ is the electrostatic potential for $L=\infty$.
We note that the boundary condition of the asymptotic solution leads to $A_t|_{\rho=\infty} = 0$.

\section{Monopole matching in Post-Newtonian order}

The leading correction to the mass and tension in the asymptotic zone are computed through 
the monopole perturbation of the post-Newtonian order, $\ord{\rho_0^{2n}}$. The near zone 
perturbation of $\ord{1/L^n}$ considered in the previous section gives the boundary condition 
in the overlap region, which behaves as $\rho_0^{2n}/L^n \rho^n$. First, we present 
the post-Newtonian equation for the gravity and the Maxwell field and then solve them. 

\subsection{Post-Newtonian order perturbation equation}

In the asymptotic zone, we consider perturbations of the metric and 
Maxwell fields to the post-Newtonian order, 
\begin{eqnarray}
  g_{\mu\nu} = \eta_{\mu\nu} + h^{(n)}_{\mu\nu} + h^{(2n)}_{\mu\nu},
   \quad
  A_\mu = A^{(n)}_\mu + A^{(2n)}_\mu,
\end{eqnarray}
where $h^{(2n)}_{\mu\nu}$ and $A^{(2n)}_\mu$ are the post-Newtonian corrections. 
In the harmonic gauge, the post-Newtonian equation becomes
\begin{eqnarray}
  - \fr{2}\Box h_{\mu\nu}^{(2n)} + 
R_{\mu\nu}^{[2]}[h^{(n)},h^{(n)}] 
=
8\pi\left[T_{\mu\nu}^{(2n)} - \frac{1}{n+1}
\eta^{\alpha\beta}T_{\alpha\beta}^{(2n)}\eta_{\mu\nu} \right]
, \label{2ndperturbein}
\end{eqnarray}
where 
we used
the Einstein equation of the Newtonian order,
$G_{\mu\nu}^{(n)}=T_{\mu\nu}^{(n)}=0$.
$R_{\mu\nu}^{[2]}$ is
the 2nd order perturbation of the Ricci tensor, which gives a source term
from the Newtonian order of Eqs.~(\ref{Newtonian-grav}) and (\ref{Newtonian-maxwell}).
The Maxwell equation in this order is given by  
\begin{equation}
 \Box A^{(2n)}_\mu + 
  \partial_\nu \left[\frac{h^{(n)}}{2}F^{(n)\nu}{}_\mu
	     - h^{(n)\nu\alpha}F^{(n)}_{\alpha\mu}
	     - h^{(n)}_{\mu\alpha}F^{(n)\nu\alpha}\right] = 0.
\label{2ndperturb_Maxwell}
\end{equation}
Using Eq.~(\ref{gauge-to-newton}), then, Eqs.~(\ref{2ndperturbein}) and
(\ref{2ndperturb_Maxwell}) become 
\begin{subequations}\label{2nd-equation}
\begin{eqnarray}
 \Box \left[h_{tt}^{(2n)} +
	   \fr{2}\left(1+\frac{\sigma^2}{2}\right)\Phi^2\right]
 = 0 ,
 \end{eqnarray}
 \begin{eqnarray}
&& 
\Box \left[ h_{ij}^{(2n)} -
\fr{2n^2}\left(1-\frac{n\sigma^2}{2}\right)\Phi^2\delta_{ij}\right]
=- \frac{n+1}{2n}\left(1-\sigma^2\right)\Phi_{,i}\Phi_{,j}+\frac{n+1}{n}(\Phi\Phi_{,i})_{,j},
\end{eqnarray}
\begin{eqnarray}
 \Box \left[A_t^{(2n)} + \fr{2n\omega_{n+1}}\frac{Q_0}{\rho_0^n}\Phi^2\right]= 0.
\end{eqnarray}\end{subequations}
The general solution is constructed by the inhomogeneous and
homogeneous solutions. The coefficients of the homogeneous solution are 
determined by the boundary condition.
We write the general solution as follows
\begin{subequations}\label{2nd-sols}
\begin{eqnarray}
 h_{tt}^{(2n)} &=& - \fr{2}\left(1+\frac{\sigma^2}{2}\right)\Phi^2 +
  s_t\frac{\rho_0^n}{L^n}\Phi,\label{2nd-sol1}\\
 h_{ij}^{(2n)} &=&
  \fr{2n^2}\left(1-\frac{n\sigma^2}{2}\right)\Phi^2\delta_{ij}+
  s_{ij}\frac{\rho_0^n}{L^n}\Phi
+ {\rm Pf}\left(\int_{[-\frac{L}{2},\frac{L}{2}]\times {\mathbf{R}}^{n+1}}  G(x,x')
{\cal S}_{ij}(x')d^{n+2}x'\right),\label{2nd-sol2}\\
 A_t^{(2n)} &=& - \frac{Q_0}{2n\omega_{n+1}\rho_0^n}\Phi^2 + s_A\frac{\rho_0^n}{L^n}\Phi ,\label{2nd-sol3}
\end{eqnarray}\end{subequations}
where ${\rm Pf}$ means the finite part of the integration and
$s_t,s_{ij},s_A$ are the dimensionless coefficients of the homogeneous
solution. 
The Green function in the compact space, $[-L/2,L/2]
\times {\mathbf R}^{n+1}$, is given by 
\begin{eqnarray}
 G(x,x') = - \fr{n\omega_{n+1}}\sum_{m=-\infty}^{\infty}\fr{(({\bf x} - {\bf x'})^2+(z-z'-mL)^2)^\frac{n}{2}}.
 \label{Greenfunc}
\end{eqnarray}
where ${\bf x}$ is the coordinate vector of ${\mathbf R}^{n+1}$, which gives $r = |{\bf x}|$.
We write the integrand as 
\begin{eqnarray}
 {\cal S}_{ij} = - \frac{n+1}{2n}\left[(1-\sigma^2)(\Phi_{,i}\Phi_{,j}) - 2(\Phi\Phi_{,i})_{,j}\right].
\end{eqnarray}

\subsection{Matching from the  near solution}

Now, we determine the coefficients, $s_t,s_{ij},s_A$ of the homogeneous term
by matching with the near solution.
The monopole moment of the near solution behaves as $\sim \rho^{-n}$.
Then the relevant terms of this order in the overlap region have the dependence of
$\rho_0^{2n}/L^n \rho^n$ in the dimensionless form. 
Transforming Eqs.~(\ref{1stsol-matched}) and (\ref{1stsol-matched-A}) into the harmonic coordinates (\ref{gauge-trans}), the near zone solution up to the relevant order becomes
\begin{subequations}\label{newton-2nd-matching}
\begin{eqnarray}
 g^{({\rm near})}_{tt} &=& - \left[ 1 - 2\lambda - (1- 2\lambda)\frac{\rho_0^n}{\rho^n}\right]+ \ord{1/L^{2n}, 1/\rho^{2n}},
 \label{newton-2nd-matching1}\\
 g^{({\rm near})}_{ij} &=& 
 \left[1 + \frac{2\lambda}{n} + \left(1 + \frac{2\lambda}{n}\right)\frac{\rho_0^n}{n\rho^n}\right]\delta_{ij}+ \ord{1/L^{2n}, 1/\rho^{2n}},
  \label{newton-2nd-matching2}\\
 A^{({\rm near})}_t &=&  \frac{2\lambda Q_0}{n\omega_{n+1}\rho_0^n} + \frac{Q_0}{n\omega_{n+1}\rho^n}(1 - \lambda)+ \ord{1/L^{2n}, 1/\rho^{2n}} .
 \label{newton-2nd-matching3}
\end{eqnarray}\end{subequations}
We extract the monopole moment of $\ord{1/L^n}$,
\begin{subequations}
\begin{eqnarray}
 g^{({\rm near})}_{tt} &\sim& -2\zeta(n)\frac{\rho_0^{2n}}{L^n\rho^n}, \\
 g^{({\rm near})}_{ij} &\sim& 
  \frac{2\zeta(n)}{n^2} \frac{\rho_0^{2n}}{L^n\rho^n}\delta_{ij}, \\
 A^{({\rm near})}_t &\sim& -\frac{\zeta(n)Q_0}{n\omega_{n+1}\rho_0^n}\frac{\rho_0^{2n}}{L^n\rho^{n}}.
\end{eqnarray}\end{subequations}
Meanwhile, the corresponding terms in the post-Newtonian solution
(\ref{2nd-sols}) are 
\begin{subequations}\label{pn-2nd-matching}
\begin{eqnarray}
 h_{tt}^{(2n)} &\sim& \left[s_t- 2\zeta(n)\left(1 +
  \frac{\sigma^2}{2}\right)\right]\frac{\rho_0^{2n}}{L^n\rho^n},\label{pn-2nd-matching1}\\
 h_{ij}^{(2n)} &\sim&
  \left[s_{ij} + \frac{2\zeta(n)}{n^2}\left(1-\frac{n\sigma^2}{2}\right)\delta_{ij}\right]\frac{\rho_0^{2n}}{L^n\rho^n},\label{pn-2nd-matching2}\\
 A_t^{(2n)} &\sim& \left[s_A - \frac{2\zeta(n)Q_0}{n\omega_{n+1}\rho_0^n}\right]\frac{\rho_0^{2n}}{L^n\rho^n}.\label{pn-2nd-matching3}
\end{eqnarray}\end{subequations}
The finite part of the integration term in Eq.~(\ref{2nd-sol2}) does not contribute to the monopole moment in the overlap region.
The matching between two solutions shows
\begin{subequations}\label{2nd-matching}
\begin{eqnarray}
 s_t &=& \sigma^2\zeta(n), \label{2nd-matching1}\\
 s_{ij} &=&  \frac{\sigma^2}{n}\zeta(n)\delta_{ij} = \frac{s_t}{n}\delta_{ij}, 
 \label{2nd-matching2}\\
 s_A &=&  \frac{\zeta(n)Q_0}{n\omega_{n+1}\rho_0^n}.\label{2nd-matching3}
\end{eqnarray}\end{subequations}
Therefore, the post-Newtonian correction is determined as
\begin{subequations}\label{2nd-sols-matched}
\begin{eqnarray}
 h_{tt}^{(2n)} &=& - \fr{2}\left(1+\frac{\sigma^2}{2}\right)\Phi^2 +
  \sigma^2 \lambda\Phi,\label{2nd-sol1-matched}\\
 h_{ij}^{(2n)} &=&
  \fr{2n^2}\left(1-\frac{n\sigma^2}{2}\right)\Phi^2\delta_{ij}+
  \frac{\sigma^2}{n}\lambda \Phi\delta_{ij} 
+ {\rm Pf}\left(\int_{[-\frac{L}{2},\frac{L}{2}]\times \mathbf{R}^{n+1}}  G(x,x')
{\cal S}_{ij}(x')d^{n+2}x'\right),\label{2nd-sol2-matched}\\
 A_t^{(2n)} &=& - \frac{Q_0}{2n\omega_{n+1}\rho_0^n}\Phi^2 + \frac{Q_0}{n\omega_{n+1}\rho_0^n}\lambda \Phi ,\label{2nd-sol3-matched}
\end{eqnarray}\end{subequations}

\section{Global charges and thermodynamics}

In this section, we compute the global charges and confirm that the first law of the thermodynamics holds 
for the current cases. 

\subsection{Global charges}

Now, we calculate the post-Newtonian correction to the global charges.
The $(n+2)$-dimensional mass $M$ and tension $\tau$ of the asymptotically Kaluza-Klein spacetime are determined by the asymptotic behavior~\cite{HO04,KSP04},
\begin{eqnarray}
 h^{({\rm asym})}_{tt} \simeq \frac{c_t}{r^{n-1}},\\
 h^{({\rm asym})}_{zz} \simeq \frac{c_z}{r^{n-1}}.
\end{eqnarray}
Then we find 
\begin{eqnarray}
 M = \frac{\omega_n L}{16\pi}(n c_t - c_z),\\
 \tau = \frac{\omega_n}{16\pi}(c_t -nc_z).
\end{eqnarray}
The total electric charge $Q$ is determined in the same way,
\begin{eqnarray}
  A_t^{({\rm asym})} \simeq \frac{Q}{(n-1)\omega_n L}\fr{r^{n-1}}.
\end{eqnarray}
To extract the global charges, we take the limit, $r \gg L,z$.
As in Eq.~(\ref{potential-asym-limit}), $\Phi$ becomes
\begin{eqnarray}
  \Phi \simeq \frac{n\omega_{n+1}}{(n-1)\omega_n} \frac{\rho_0^n}{Lr^{n-1}},
\end{eqnarray}
Since the Green function in Eq.~(\ref{Greenfunc}) has the similar behavior as $\Phi$, the integration becomes
\begin{eqnarray}
  \int_{[-\frac{L}{2},\frac{L}{2}]\times \mathbf{R}^{n+1}}  
G(x,x'){\cal S}_{ij}(x')d^{n+2}x' \simeq - \fr{(n-1)\omega_n}\fr{Lr^{n-1}} 
\int_{[-\frac{L}{2},\frac{L}{2}]\times \mathbf{R}^{n+1}} {\cal S}_{ij}(x')d^{n+2}x'
\end{eqnarray}
Then, Eq.~(\ref{2nd-sols-matched}) gives
\begin{eqnarray}
 h_{tt}^{(\mathrm{asym},2n)} &\simeq& \frac{n\omega_{n+1}\sigma^2\zeta(n)}{(n-1)\omega_n}\frac{\rho_0^{2n}}{L^{n+1}}\fr{r^{n-1}},\label{asym-tt}\\
 h_{zz}^{(\mathrm{asym},2n)} &\simeq& \frac{\omega_{n+1}\sigma^2\zeta(n)}{(n-1)\omega_n}\frac{\rho_0^{2n}}{L^{n+1}}\fr{r^{n-1}}
+
  \frac{(n+1)(1-\sigma^2)}{2n(n-1)\omega_nLr^{n-1}}{\rm Pf}
  \left(\int d^{n+1}\mathbf{x} \int^{L/2}_{-L/2}\!dz \Phi_{,z}^2\right). \label{asym-zz}
\end{eqnarray}
The second term in ${\cal S}_{zz}(x)$ does not contribute to
Eq.~(\ref{asym-zz}) because it is the total derivative.
The finite part of the integration is computed in Ref.~\cite{GK05}
\begin{eqnarray}
  {\rm Pf}\left(\int\!d^{n+1}\mathbf{x}\int^{L/2}_{-L/2}\!dz\Phi_{,z}^2 \right)= - \frac{(n-1)n\zeta(n)\omega_{n+1}\rho_0^{2n}}{L^n}.
\end{eqnarray}
As a result, the ADM mass and tension become
\begin{eqnarray}
 M &=& \frac{(n+1)\omega_{n+1}\rho_0^n}{16\pi}
 \left(1+ \fr{2}(1+\sigma^2)\lambda\right)\nonumber\\
  &=& M_0 \left(1 + \fr{2}(1+\sigma^2)\lambda \right)
 \label{adm-mass},\\
 \tau L &=& \frac{n(n+1)\omega_{n+1}\rho_0^n}{32\pi}(1-\sigma^2)\lambda\nonumber\\
  &=& \frac{n}{2}M_0(1-\sigma^2)\lambda. \label{tension}
\end{eqnarray}
From Eq.~(\ref{2nd-sol3-matched}), the total charge $Q$ becomes
\begin{eqnarray}
 Q = Q_0(1 +\lambda).
\end{eqnarray}
At $\sigma=0$, these results reproduce the results in the neutral case~\cite{GK05}.

Moreover, in the extremal case ($\sigma=1$) we can confirm that our
results correspond to the exact solutions in Ref.~\cite{Myers:1986rx}.
The exact solutions of the ($n+3$)-dimensional extremal caged black hole
are given by 
\begin{equation}
 ds^2 = - H^{-2} dt^2 + H^{2/n} \delta_{ij} dx^i dx^j, \quad
  A_\mu dx^\mu = - \sqrt{\frac{n+1}{8\pi n}}H^{-1} dt,
\end{equation}
where 
\begin{equation}
 H(x^i) = 1 + \mu \sum_{k=-\infty}^{\infty}
  \frac{1}{(r^2 + (z-kL)^2)^{\frac{n}{2}}}.
\end{equation}
If we set $\mu = \rho_0^n (1 + \lambda)/2$, at $\sigma=1$ the
perturbative solutions constructed agree with the above
solutions up to $\mathcal O(\lambda)$.

There is the apparent difference between our results and the results by the EFT calculation~\cite{GRS09}. This 
simply comes from the ambiguity of the parameterization. For example, see Eqs. (3.7) and (3.25) in Ref. \cite{GRS09} 
which are $M$ and $\tau$. 
Changing the gauge condition in Eq.~(\ref{fix-gauge-ambiguity}) gives the different length scale $\rho_0 \rightarrow \rho_0(1 + \epsilon_1(\sigma))$ and charge parameterization $\sigma \rightarrow \sigma(1 + \epsilon_2(\lambda,\sigma))$.
We note that, to keep $\sigma=1$ to be the extreme limit, one requires $\epsilon_2(\lambda,1)=0$ further. 
To reproduce the results by the EFT, we should impose the following gauge condition,
instead of Eq.~(\ref{fix-gauge-ambiguity}), 
\begin{eqnarray}
  C_2 = - \frac{2\rho_+^nC_1}{n\rho_0^n}, C_3=C_1, \alpha_0=0 \label{another-gauge}.
\end{eqnarray}

\subsection{First law and the Smarr formula}

It is ready to confirm that our solutions satisfy the Smarr formula
\begin{eqnarray}
  nM = \frac{n+1}{8\pi}\kappa \mathcal{A} + \tau L+  n QU, \label{smarreq}
\end{eqnarray}
and the first law
\begin{eqnarray}
 dM = \tau dL + \fr{8\pi} \kappa d\mathcal{A} + UdQ.\label{1stlaw}
\end{eqnarray}
Collecting the thermodynamic variables from
Eqs.~(\ref{surface-gravity}), (\ref{area}), 
(\ref{electrostatic-potential}), (\ref{adm-mass}) and (\ref{tension}), 
we write
\begin{equation}
\begin{aligned}
 M = M_0 \left(1 + \frac{1}{2}(1+\sigma^2)\lambda\right), \quad
 \mathcal{A} = \mathcal{A}_0 \left(1 + \frac{n+1}{n}\lambda\right),
 \quad
 Q = Q_0 (1 + \lambda), \\
 \kappa = \kappa_0 \left(1 - \frac{n+1}{n}\lambda\right), \quad
 U = U_0 (1 + \sqrt{1-\sigma^2}\lambda), \quad
 \tau L = \frac{n}{2}M_0(1-\sigma^2)\lambda,
\end{aligned}
\end{equation}
where 
\begin{eqnarray}
 &M_0& = \frac{(n+1)\omega_{n+1}\rho^n_0}{16\pi}, \quad
  \mathcal{A}_0 = \omega_{n+1}\rho^{n+1}_0 
  \left(\frac{1+\sqrt{1-\sigma^2}}{2}\right)^{\frac{n+1}{n}}, \quad
  Q_0 = \sqrt{\frac{n(n+1)}{32\pi}}\omega_{n+1}\rho_0^n \sigma \nonumber\\
  &\kappa_0&=\frac{n\rho_0^n}{2\rho_+^{n+1}}\sqrt{1-\sigma^2}, \quad U_0 = \frac{Q_0}{n\omega_{n+1}\rho_+^n}.
\end{eqnarray}
From the direct computation, 
the above expressions are easily confirmed to follow Eq.~(\ref{smarreq}) up to $\ord{\lambda}$. 

To confirm the first law for the current case, we take the variation of $M$ with $L$, ${\mathcal A}$ and $Q$.
Changing the variables from $(\rho_0, \lambda, \sigma)$ to $(L, {\mathcal A}, Q)$,
the variation becomes
\begin{eqnarray}
\left[ \begin{array}{c} (\partial_L M)_{\mathcal{A}, Q} \\
       (\partial_{\mathcal A} M)_{L, Q}  \\ 
	(\partial_Q M)_{L, \mathcal{A}} \end{array}\right]
 =  \frac{\partial(\rho_0,\lambda,\sigma)}{\partial(L, {\cal A}, Q)}
 \left[ \begin{array}{c} (\partial_{\rho_0} M)_{\lambda, \sigma} \\ 
	(\partial_\lambda M)_{\rho_0, \sigma} \\ 
	(\partial_\sigma M)_{\rho_0, \lambda} \end{array}\right]
 =  \frac{\partial(\rho_0,\lambda,\sigma)}{\partial(L, {\cal A}, Q)}
 \left[ \begin{array}{c} n M/\rho_0 \\ 
	M_0(1+\sigma^2)/2 \\ 
	M_0\sigma\lambda \end{array}\right].
\end{eqnarray}
Using $L = \zeta(n)^{\fr{n}}\rho_0\lambda^{-\fr{n}}$, 
the Jacobian $\partial(\rho_0,\lambda,\sigma)/\partial(L, {\cal A}, Q)$ is computed as 
\begin{eqnarray}
\frac{\partial(\rho_0,\lambda,\sigma)}{\partial(L,{\cal A}, Q)} &=& \left( \frac{\partial(L,{\cal A},Q)}{\partial(\rho_0,\lambda, \sigma)}\right)^{-1} \nonumber\\
 &=&\left[ \begin{array}{ccc}
  \rho_0\lambda/L &-n\lambda /L & 0\\
  \rho_0\kappa_0(1-(2n+1)\lambda/n)/8\pi nM_0&\lambda \kappa_0 /8\pi M_0 &- \sigma\kappa_0 /8\pi M_0\\
  \rho_0 U_0(1-2\lambda)/nM_0&\lambda U_0/M_0 & U_0(1+\sqrt{1-\sigma^2})\sqrt{1-\sigma^2}/\sigma M_0
  \end{array}\right].
\end{eqnarray}
Using this, we can show that 
the variation of $M$ satisfies the first law~(\ref{1stlaw}) up to $\ord{\lambda}$
\begin{eqnarray}
\left[ \begin{array}{c} (\partial_L M)_{\mathcal{A}, Q} \\
       (\partial_{\cal A} M)_{L, Q} \\
       (\partial_Q M)_{L, \mathcal{A}} \end{array}\right]
 = \left[ \begin{array}{c} \tau \\ \kappa/8\pi \\ U \end{array}\right].
\end{eqnarray}

Note that this is also achieved through the Harrison transformation from the neutral seed solutions
~\cite{KKRS09}. On the other hand, our argument gives us a direct confirmation using the direct 
construction of the perturbative solutions.

\section{Summary and discussion}

In this paper, we constructed the perturbative solution of
 the small black holes with the Maxwell charge
in the caged spacetime using the matched asymptotic expansion. 
The expansion of the Maxwell field can be performed in the same way as the metric. 
Although our results seem different with the EFT calculation in Ref.~\cite{GRS09},
this is just the difference in the parameterization. We also confirmed the first law and the 
Smarr formula, which were shown for a sequence of charged black objects in Kaluza-Klein 
spacetime in Ref.~\cite{KKRS09}.

The another way to construct the charged solution
 is using of the Harrison transformation~\cite{KKRS09,GR98}
which produces the charged dilatonic solution from the neutral seed solution. 
However, the charged stationary solution such like the rotating black ring cannot be 
produced by the Harrison transformation. In such case, one should rely on some perturbative 
methods with the Maxwell field as we used in this paper. 
Since the charged black ring is studied only by the blackfold approach~\cite{CEP11},
it is also interesting to study the charged black ring beyond the blackfold approach.

We have not included the finite size effect which comes from the deformation of the horizon.
To see this effect, we must consider the multipole perturbation which will be studied in the future work.

\begin{acknowledgments}
We would like to thank Kentaro~Tanabe and Seiju~Ohashi for their useful discussions. 
RS thanks Prof. Takashi Nakamura for his continuous encouragement. 
RS is supported by the Grant-in-Aid for the Global COE Program ``The 
Next Generation of Physics, Spun from Universality 
and Emergence'' from the Ministry of Education, Culture, Sports, Science
and Technology (MEXT) of Japan.
SK is partially supported by the JSPS Strategic Young Researcher Overseas
Visits Program for Accelerating Brain Circulation ``Deeping and Evolution of Mathematics
and Physics, Building of International Network Hub based on OCAMI''.  TS is supported by
Grant-Aid for Scientific Research from the Ministry of Education, Culture, Sports, Science,
and Technology (MEXT) of Japan (Nos.~21244033 and 19GS0219). 
\end{acknowledgments}

\appendix
\section{Gegenbauer polynomials}
\label{app:Gegenbauer}

The Gegenbauer polynomials $C_l^{\nu}(x)$ are defined as the coefficients
of the following generating function,
\begin{eqnarray}
 \fr{(1 - 2xt + t^2)^\nu} = \sum_{l=0}^\infty C_l^\nu (x)t^l,
\end{eqnarray}
where $C^\nu_l(x)$ is written as
\begin{eqnarray}
 C_l^\nu(x) = \frac{(-1)^l}{l!2^l}\frac{\Gamma(\nu+\fr{2})\Gamma(l+2\nu)}{\Gamma(2\nu)\Gamma(l+\nu+\fr{2})}(1-x^2)^\frac{1-2\nu}{2} 
\frac{d^l}{dx^l}[(1-x^2)^\frac{2l+2\nu-1}{2}].
\end{eqnarray}
It follows the equation
\begin{eqnarray}
 (1-x^2)y'' - (2\nu+1)y' + l(l+2\nu)y = 0.
\end{eqnarray}

\section{Asymptotic limit of the Newtonian potential}

In the limit, $r \gg z,L$, the Newtonian potential $\Phi$ in Eq.~(\ref{Newtonian-grav1})
behaves as
\begin{eqnarray}
  \Phi &=& \sum_{m=1}^\infty\frac{\rho_0^n}{(r^2+m^2L^2)^\frac{n}{2}}\left[\left(1-\frac{2mLz}{r^2+m^2L^2} + \frac{z^2}{r^2+m^2L^2}\right)^{-\frac{n}{2}}
  +\left(1+\frac{2mLz}{r^2+m^2L^2} + \frac{z^2}{r^2+m^2L^2}\right)^{-\frac{n}{2}}\right] + \frac{\rho_0^n}{(r^2+z^2)^\frac{n}{2}}\nonumber\\
   &=& \frac{2\rho_0^n}{Lr^{n-1}}\sum_{m=1}^\infty \frac{L/r}{(1 + (mL/r)^2)^\frac{n}{2}}  + \ord{1/r^n} \nonumber\\
   & = & \frac{2\rho_0^n}{Lr^{n-1}}\int_0^\infty \frac{dt}{(1+t^2)^\frac{n}{2}} + \ord{1/r^n} 
   \nonumber \\
   & = & \frac{n\omega_{n+1}}{(n-1)\omega_n} \frac{\rho_0^n}{Lr^{n-1}} + \ord{1/r^n}. \label{potential-asym-limit}
\end{eqnarray}

\section{Multipole perturbation in the near zone}
In this section, we give the equation for the multipole perturbation around the charged black hole.
\subsection{Ansatz}
The static spacetime can be written in this form,
\begin{eqnarray}
ds^2 = g_{tt}(x)dt^2 + g_{ij}(x)dx^i dx^j.
 \label{static-coord}
\end{eqnarray}
In this ansatz, we consider the linear perturbations around the black hole as  
$ g_{\mu\nu} = g^{({\rm BH})}_{\mu\nu} + h_{\mu\nu}$, 
where the background metric is given by 
\begin{equation}
 g^{({\rm BH})}_{\mu\nu} dx^\mu dx^\nu
  = - f(\rho)dt^2 + \frac{d\rho^2}{f(\rho)} + \rho^2 d\Omega^2_{n+1}.
\end{equation}
Since the background metric has $\mathrm{SO}(n+2)$ symmetry, we can expand the
perturbations in terms of the spherical harmonics on the $(n+1)$-dimensional sphere.
Here, we consider only scalar perturbations.
We explicitly write all nonzero components of the perturbations,
\begin{subequations}\label{multi-perturb}
\begin{eqnarray}
 h_{tt} &=& - f \sum_{l=0}^{\infty} A_l(\rho)\mathsf Y_{l},\label{multi-perturb1}\\
 h_{\rho\rho}  &=& f^{-1}\sum_{l=0}^{\infty} B_l(\rho)\mathsf Y_{l},\label{multi-perturb2}\\
 h_{\rho a}  &=& f^{-1}\sum_{l=1}^{\infty} C_l(\rho)\mathcal{D}_a \mathsf Y_{l},\label{multi-perturb3}\\
 h_{ab} &=& \rho^2 \sum_{l=1}^{\infty}D_l(\rho)
  [\mathcal D_a \mathcal D_b - \frac{1}{n+1}\gamma_{ab}\mathcal D^2]
  \mathsf Y_{l}+ \rho^2 \gamma_{ab} \sum_{l=0}^\infty E_l(\rho) \mathsf Y_{l},\label{multi-perturb4}
\end{eqnarray}\end{subequations}
where $\gamma_{ab}$ is the metric of the $(n+1)$-dimensional unit sphere
and $\mathcal D_a$ denotes the covariant derivative with respect to
$\gamma_{ab}$.
Here, $\mathsf Y_l$ is the spherical harmonics with $l$-th multipole
moment, which satisfies 
$\mathcal D^2 \mathsf Y_l = - l(l+n)\mathsf Y_l$.
The gauge transformation for the metric perturbations is given by 
$h_{\mu\nu} \rightarrow h_{\mu\nu} + \mathcal L_\xi g_{\mu\nu}^{(\mathrm{BH})}$ under the infinitesimal coordinate
transformations $x^\mu \rightarrow x^\mu - \xi^\mu$.
Then, the components transform as 
\begin{equation}
\begin{aligned}
 A_l \rightarrow A_l + \frac{f'}{f}\xi^\rho_l, \quad 
  B_l \rightarrow B_l + 2 {\xi^\rho_l}' - \frac{f'}{f}\xi^\rho_l, \quad
  C_l \rightarrow C_l + \xi^\rho_l + f\rho^2 \zeta_l,\\
 D_l \rightarrow D_l + 2 \zeta_l, \quad
 E_l \rightarrow E_l + \frac{2}{\rho}\xi^\rho_l
 - \frac{2l(l+n)}{n+1} \zeta_l,
\end{aligned}
\end{equation}
where we have expanded $\xi^\mu$ with the harmonics $\mathsf Y_l$ 
and $\zeta_l$ is defined as  
$\xi^a = \zeta_l \mathcal D^a \mathsf{Y}_l$.
Note that we have not transformed the time coordinate $t$. 
Using the above gauge degrees of freedom, $\xi^\rho_l$ and $\zeta_l$, we
can set $C_l = D_l = 0$~\cite{GK04}. In the monopole case, $C_0$ and
$D_0$ are automatically zero and then $\xi^\rho_0$ becomes the residual gauge.
For the vector potential, we write the perturbations, $\delta A_t$, 
$\delta A_\rho$ and $\delta A_a$ as
\begin{subequations}\label{multi-gauge}
\begin{eqnarray}
 \delta A_t &=& \sum_{l=0}^\infty a_l(\rho)\mathsf Y_{l},\\
 \delta A_\rho &=& \sum_{l=0}^\infty b_l(\rho)\mathsf Y_{l},\\
 \delta A_a &=& \sum_{l=1}^\infty c_l(\rho)\mathcal D_a \mathsf Y_{l}.
\end{eqnarray}\end{subequations}
The $U(1)$ gauge transformation of $\delta A_\mu \rightarrow \delta A_\mu + \partial_\mu\psi$, where $\psi$ is a scalar function, transforms the potential as
\begin{eqnarray}
 b_l \rightarrow b_l + \psi_l', \quad c_l \rightarrow  c_l + \psi_l
\end{eqnarray}
where $\psi_l(\rho)$ is the expansion coefficient of $\psi$ for 
the spherical harmonics $\mathsf Y_l$.
Therefore, we can set $c_l=0$ ($l \neq 0$) by choosing $\psi_l$ appropriately. 
In the monopole case, we also set $b_0 = 0$, because $c_0$ is automatically zero.
The field strength becomes
\begin{subequations}\label{field-strength}
\begin{eqnarray}
  \delta F_{t\rho} = -\sum_{l=0}^\infty a_l' \mathsf{Y}_l,\label{field-strength1}\\
  \delta F_{t a} = -\sum_{l=1}^\infty a_l \mathcal D_a \mathsf{Y}_l,\label{field-strength2}\\
 \delta F_{\rho a} = -\sum_{l=1}^\infty b_l \mathcal D_a \mathsf{Y}_l.\label{field-strength3} 
\end{eqnarray}\end{subequations}

\subsection{Perturbation equation}

The Einstein equation for the linear perturbation is
\begin{eqnarray}
 \delta R_{\mu\nu} &=& 8\pi
  \left[\delta T_{\mu\nu}-\fr{n+1}
(g^{(\mathrm{BH})\alpha\beta}\delta T_{\alpha\beta} - T^{(0)}{}^{\alpha\beta}h_{\alpha\beta})g^{(\mathrm{BH})}_{\mu\nu}-\fr{n+1}T^{(0)} h_{\mu\nu}\right]\nonumber\\
  &\equiv&  8\pi \delta S_{\mu\nu}
\end{eqnarray}
Under the metric ansatz~(\ref{multi-perturb}), the non-vanishing components
of $\delta R_{\mu\nu}$ become 
\begin{subequations}\label{multi-ricci}
\begin{eqnarray}
\delta R_{tt} &=& - \frac{f}{4\rho^2}[ - 2\rho^2fA_l'' - (3\rho^2f'+2(n+1)\rho f)A_l' + \rho^2f'B_l'\nonumber\\
 & & - (n+1)\rho^2f'E_l' - 2(\rho^2f''+(n+1)\rho f')(A_l-B_l) \nonumber\\
 & & + 2l(l+n)A_l]\mathsf Y_l\label{ricci-tt}
 \end{eqnarray}
 \begin{eqnarray}
  \delta R_{\rho\rho} &=& \frac{1}{4\rho^2 f}
 [- 2\rho^2fA_l''-2(n+1)\rho^2fE_l''-3\rho^2f'A_l' \nonumber\\
 & & +(\rho^2f'+2(n+1)\rho f)B_l' - (n+1)(\rho^2f'+4\rho f)E_l'\nonumber\\
 & & + 2l(l+n)B_l]\mathsf Y_l\label{ricci-rhorho}
 \end{eqnarray}
 \begin{eqnarray}
  \delta R_{\rho a}&=&\frac{1}{4\rho f}[ - 2\rho fA_l'-2n\rho fE_l'-(\rho
 f'-2f)A_l\nonumber\\
 & &+ (\rho f'+2nf)B_l]\mathcal D_a \mathsf Y_l \label{ricci-rhochi}
 \end{eqnarray}
 \begin{eqnarray}
  \delta R_{ab}  &=&  \fr{2}[-\rho^2fE_l'' - (\rho^2f'+2(n+1)\rho f)E_l'-\rho f(A_l'-B_l') 
\nonumber\\
  & &+ 2(\rho f'+nf)(B_l-E_l)+  l(l+n)E_l]\mathsf{Y}_l \gamma_{ab}\nonumber\\
  & &- \fr{2}[A_l+B_l+(n-1)E_l] \mathcal D_a \mathcal D_b \mathsf{Y}_l\label{ricci-ab}
 \end{eqnarray}
\end{subequations}
where we write only the $l$-th pole components. 
On the other hand, the non-vanishing components of $\delta S_{\mu\nu}$ becomes
\begin{subequations}\label{s-tensor}
\begin{eqnarray}
  \delta S_{tt} &=& -f\frac{n}{n+1}(2\cE a_l' + \cE^2 B_l)\mathsf Y_l,\label{s-tensor1}\\
 \delta S_{\rho\rho} &=& \frac{1}{f}\frac{n}{n+1}
  (2\cE a_l' + \cE^2 A_l)\mathsf Y_l,\label{s-tensor2}\\
 \delta S_{ab} &=& \rho^2 \gamma_{ab}
  \frac{1}{n+1}[-2\cE a_l' + \cE^2 (E_l - A_l - B_l)]
  \mathsf Y_l,\label{s-tensor3}\\
 \delta S_{\rho a} &=& \frac{1}{f}\cE a_l \mathcal D_a\mathsf Y_l, \label{s-tensor4}\\
 \delta S_{ta} &=& f \cE b_l \mathcal D_a\mathsf Y_l, \label{s-tensor5}
\end{eqnarray}\end{subequations}
where $\cE$ is defined by Eq.~(\ref{gauge-field-bg}). From $\delta R_{ta} = 0$ and Eq.~(\ref{s-tensor5}),
we see $b_l=0$ for $l\neq 0$.
Moreover, $a_l \, (l \neq 0)$ is determined by using the Einstein equation from
Eqs.~(\ref{ricci-rhochi}) and (\ref{s-tensor4}). Therefore, we do not need to solve the Maxwell equation directly 
to determine $a_l\,(l \neq 0)$. 

\subsection{Master Equations}

From the traceless part of Eqs.~(\ref{ricci-ab}) and
(\ref{s-tensor3}), we obtain the following algebraic relation 
\begin{eqnarray}
 A_l+B_l+(n-1)E_l=0, \label{eq:constraint}
\end{eqnarray}
because 
$[(n+1)\mathcal D_a\mathcal D_b - \gamma_{ab}\mathcal D^2]\mathsf Y_l \neq 0$ when $l > 1$.
Note that for $l=0, 1$ we do not have any constraint from the perturbation
equations but we can require this equation by using the residual gauge (See Eq. (\ref{monopole-res-gauge})).

Now, we introduce the following master variables,
\begin{eqnarray}
 A_l \equiv \Psi - n\Pi , \; B_l \equiv \Pi -\Psi, \; E_l \equiv \Pi, \label{master-variables}
\end{eqnarray}
which satisfy Eq.~(\ref{eq:constraint}).
Using the master variables, we obtain the following master equations from the
perturbation equations 
\begin{subequations}\label{master-equations1}
\begin{eqnarray}
 (n+1)\rho^2 f\Pi'' + (n+1)\rho f\Pi' -(n+1)l(l+n)\Pi  = -2l(l+n)\Psi,~\label{master-equations1a}
 \end{eqnarray}
 \begin{eqnarray}
\rho^2f\Psi'' + (2\rho^2f'+(3n+17)\rho f)\Psi'
+(2n^2-l(l+n))\Psi=0. \label{master-equations1b}
\end{eqnarray}\end{subequations}
Using the new dimensionless variable $x\equiv (\rho/\rho_0)^{n}$,
Eqs.~(\ref{master-equations1}) become
\begin{subequations}\label{master-equations2}
\begin{eqnarray}
 fx^2\frac{d^2\Pi}{dx^2} + 2fx\frac{d\Pi}{dx} -
  \pld(\pld+1)\Pi=-\frac{2}{n+1}\pld(\pld+1)\Psi \label{master-equations2a}
\end{eqnarray}
and
\begin{eqnarray}
fx^2\frac{d^2\Psi}{dx^2} + (2x\frac{df}{dx}+4f)x\frac{d\Psi}{dx}+(2-\pld(\pld+1))\Psi=0\nonumber\\
\label{master-equations2b}
\end{eqnarray}\end{subequations}
where $\pld\equiv l/n$. Using $f(x)=(x-x_+)(x-x_-)/x^2$, Eq.~(\ref{master-equations2b}) becomes
\begin{eqnarray}
 (x-x_+)(x-x_-)\frac{d^2\Psi}{dx^2} + (4x-2x_+-2x_-)\frac{d\Psi}{dx} +(2-\pld(\pld+1))\Psi=0\label{master-equations2b-2}
\end{eqnarray}
This equation has the three singular points : $x=x_-,x_+$ and $\infty$.
 Therefore, the equation for $\Psi$ becomes the hypergeometric differential equation. Using $\xi =
(x-x_-)/(x_+-x_-)$, it becomes standard form,
\begin{eqnarray}
 \xi(1-\xi)\frac{d^2\Psi}{d\xi^2}+(2-4\xi)\frac{d\Psi}{d\xi}+(\pld+2)(\pld-1)\Psi=0.\nonumber\\
  \label{hypergeom1}
\end{eqnarray}
This is the same with one for the neutral case~\cite{GK04}. Similarly, defining $\tilde{\Pi}\equiv x\Pi/(x_+-x_-)$,
Eq.~(\ref{master-equations2a}) becomes the hypergeometric differential equation with a source term
\begin{eqnarray}
 \xi(1-\xi)\frac{d^2\tilde{\Pi}}{d\xi^2} +
  \pld(\pld+1)\tilde{\Pi}=\frac{2\pld(\pld+1)}{n+1}(\xi+e)\Psi, \nonumber\\
  \label{hypergeom2}
\end{eqnarray}
where we defined $e \equiv x_-/(x_+-x_-)$.
So, our multipole solution, in general, can be written by four hypergeometric functions and the characteristic solution for Eq.~(\ref{hypergeom2}).
In the neutral case, we do not need to solve the equation for $\Pi$, because $\Pi$ is obtained directly from $\Psi$ using Eqs.~(\ref{ricci-rhochi}) and $\delta R_{\rho a}=0$.

\section{Gauge transformation into the harmonic coordinates}

To match the near solution with the asymptotic solution in the overlap
region, we need the near solution written in the harmonic gauge.
If we write the harmonic coordinates as $x^\mu_h(x^\mu) =
(t,\rho_h(\rho),\chi,\dots)$, the harmonic condition $(\nabla^2 x^\mu = 0)$ becomes
\begin{eqnarray}
 \nabla^2(\rho_h \cos\chi)=0. \label{harmonic-cond}
\end{eqnarray}
In the above, $\nabla^2$ is the Laplacian for the full near metric $g_{\mu\nu}
= g^{(\mathrm{BH})}_{\mu\nu} + h_{\mu\nu}$.
Since the monopole perturbation does not depend on $\chi$,
Eq.~(\ref{harmonic-cond}) up to $\ord{1/L^n}$ becomes
\begin{eqnarray}
 \left(1-\fr{2}h^{(n)}\right)\partial_\rho\left[\left(1+\fr{2}h^{(n)}\right)\rho^{n+1}f(1-B_0)\partial_\rho\rho_h\right]
= (n+1)(1+E_0)\rho^{n-1}\rho_h, 
\end{eqnarray}
where $h^{(n)} = A_0 + B_0 + (n+1)E_0$ is the trace of $h_{\mu\nu}^{(n)}$.

We expand $\rho_h$ as $\rho_h = \rho_h^{(0)} + \rho_h^{(n)}$, where $\rho_h^{(n)}$ is of $\ord{1/L^n}$.
Then, the $\ord{1/L^0}$ equation becomes
\begin{eqnarray}
 \partial_\rho(\rho^{n+1}f\partial_\rho\rho_h^{(0)})= (n+1)\rho^{n-1}\rho_h^{(0)}.
\end{eqnarray}
Assuming $\rho_h \rightarrow \rho$ as $\rho \rightarrow \infty$, the solution is given by
\begin{eqnarray}
 \rho_h^{(0)} = \rho - \fr{2n}\frac{\rho_0^n}{\rho^{n-1}} + \ord{\rho_0^{2n}/\rho^{2n-1}}.
\end{eqnarray}
The $\ord{1/L^n}$ equation is
\begin{eqnarray}
 \partial_\rho(\rho^{n+1}f\partial_\rho\rho_h^{(n)})
 - (n+1)\rho^{n-1}\rho_h^{(n)}
 = \partial_\rho (\rho^{n+1}fB_0 \partial_\rho \rho_h^{(0)})\nonumber\\
  -
 \fr{2}\rho^{n+1}\partial_\rho h^{(n)} f \partial_\rho \rho_h^{(0)} - (n+1)\rho^{n-1}E_0\rho_h^{(0)}.\;\;
\end{eqnarray}
 Note that the $\ord{1/L^n}$ equation is trivial in the gauge choice~(\ref{fix-gauge-ambiguity}). 
Here we consider cases for general $C_2,C_3,\alpha_0$. 
 Assuming $\rho_h^{(n)} \rightarrow 0$ as $\rho\rightarrow \infty$, the solution becomes
\begin{eqnarray}
 \rho_h^{(n)} = \frac{C_3+nC_2}{2n}\frac{\rho_0^n}{\rho^{n-1}} + \ord{\rho_0^{2n}/\rho^{2n-1}}.
\end{eqnarray}
As a result, we have
\begin{eqnarray}
 \rho_h(\rho) = \rho - \frac{\rho_0^n}{2n\rho^{n-1}}( 1 - nC_2 - C_3) + \ord{\rho_0^{2n}/\rho^{2n-1}}.
\label{gauge-trans}
\end{eqnarray}
This equation gives the coordinate transformation from $(t,\rho,\chi,\dots)$ into $(t,\rho_h(\rho),\chi,\dots)$.
In the harmonic coordinates, the near zone metric and the gauge field become
\begin{subequations}\label{gauge-transformed}
\begin{eqnarray}
  g_{tt}^{({\rm near})} &=& -f (\rho)(1 + A_0(\rho))  = - \left( 1 - 2C_1 - \left(1 - 2C_1 - C_3\right)\frac{\rho_0^n}{\rho_h^n}\right) + \ord{1/L^{2n}, 1/\rho_h^{2n}},\\
  g_{\rho_h\rho_h}^{({\rm near})} &=& \left(\dell{\rho}{\rho_h}\right)^2 f^{-1}(\rho)(1 + B_0(\rho)) = 1 + \frac{2C_1}{n} + \frac{\rho_0^n}{n\rho_h^n} \left(1+ \frac{2C_1}{n} - C_3\right)+ \ord{1/L^{2n}, 1/\rho_h^{2n}},\\
  g_{\chi\chi}^{({\rm near})} &=& \rho^2(1+E_0(\rho)) = \rho_h^2\left(1 + \frac{2C_1}{n} +  \frac{\rho_0^n}{n\rho_h^n}\left(1+ \frac{2C_1}{n}-C_3\right) + \ord{1/L^{2n}, 1/\rho_h^{2n}}\right),\\
  F_{t\rho_h}^{({\rm near})} &=&  \frac{Q_0}{\omega_{n+1}\rho_h^{n+1}}(1- \alpha_0 - 2C_1 +  \ord{1/L^{2n}, 1/\rho_h^{n}}).
\end{eqnarray}\end{subequations}


\end{document}